\documentclass[twocolumn,showpacs]{revtex4}
\usepackage[xdvi]{graphicx}%
\usepackage{dcolumn}
\usepackage{epsfig}

\newcommand{\ret}{\nonumber\\}

\newcommand{\bkt}[1]{\left\langle#1\right\rangle}
\newcommand{\sbkt}[1]{\langle#1\rangle}

\newcommand{\bsr}{\mathbf{r}}
\newcommand{\bst}{\mathbf{t}}
\newcommand{\bsR}{\mathbf{R}}
\newcommand{\bsp}{\mathbf{p}}
\newcommand{\bsP}{\mathbf{P}}
\newcommand{\eqref}[1]{(\ref{#1})}

\newcommand{\tilnu}{\tilde{\nu}}
\newcommand{\tilmu}{\tilde{\mu}}
\newcommand{\PRX}{\bsP,\bsR,X}
\newcommand{\PR}{\bsP,\bsR}
\newcommand{\PRXp}{\bsP',\bsR',X'}
\newcommand{\PRp}{\bsP',\bsR'}
\newcommand{\calC}{\mathcal{C}}
\newcommand{\calR}{\mathcal{R}}

\begin{document}
\title{Second law of thermodynamics for macroscopic mechanics coupled to thermodynamic degrees of freedom}
\author{Christian Maes${}^{1}$ and Hal Tasaki${}^{2}$}
\affiliation {${}^1$Instituut voor Theoretische Fysica,
K.U.Leuven, Celestijnenlaan 200D, B-3001, Belgium
\\
${}^2$Department of Physics, Gakushuin University,
Mejiro, Toshima-ku, Tokyo 171-8588, Japan}

\date{\today}

\vspace{5in}

\begin{abstract}
Based only on classical Hamiltonian dynamics, we prove
the maximum work principle in a system where macroscopic dynamical
degrees of freedom are intrinsically coupled to 
microscopic degrees of freedom.  Unlike recent identities
between irreversible work and free energy, such as in the
Jarzynski relation, the macroscopic dynamics is not governed by an
external action but undergoes the back reaction of the microscopic
degrees of freedom. 
Our theorems cover such physical situations
as impact between macroscopic bodies, thermodynamic machines, and
molecular motors.
\end{abstract}


\maketitle


The second law of thermodynamics is one of the most important
principles in physics.  Since its initial conception various
formulations have coexisted and corresponding derivations from
microscopic mechanics have been argued (and criticized).  Most
valuable in that respect remains the recognition by the founding
fathers Maxwell, Boltzmann, Gibbs and Einstein of the importance
of the huge separation between microscopic and macroscopic scales.
The second law indeed realizes a bridge between the microscopic
physics
laws and the macroscopic reality.

In the present letter we consider a mechanical system containing a
huge number of degrees of freedom.  The macroscopic motion of the
system is coupled directly and generally with these microscopic
degrees. We focus on the second law in its maximum work
formulation  \cite{en:work}, getting an upper bound for the change in energy of
the macroscopic degrees of freedom in terms of the change in free
energy.  To our knowledge the present rigorous derivation starting
from Hamiltonian dynamics is unique in that respect.
An advantage of our formulation is that  we deal only with macroscopic energy transfer, whose mechanical definition (unlike that of entropy) is manifest.

Our work has to be contrasted with the results in \cite{en:second}
concerning e.g. irreversible work - free energy relations, where
one always treats the dynamics of the macroscopic coordinates as
operated upon by an {\it external} agent. The operation is 
there implemented as time-dependent Hamiltonian, whose
time-dependence follows a protocol which is fixed prior to the
operation. In other words, the external agent is never affected by
back-reaction from the system under operation.
Such an idealization is justified if, for example, a gas in a
cylinder is compressed by moving an infinitely heavy rigid piston
whose motion or shape is never affected by the behavior of the
gas. In reality, however, any part (including the piston) of a
physical system obeys its equation of motion, and is also affected
by the motion of other constituents of the same system. It is
therefore desirable (if only from a theoretical point of view) to
treat macroscopic degrees of freedom as truly dynamical degrees of
freedom.

There are however much more delicate situations where macroscopic
(or mesoscopic) motions  are coupled intrinsically with
microscopic degrees of freedom and where our treatment becomes
very relevant. Examples include impact between several macroscopic
bodies, various thermodynamic machines, and biological molecular
motors. In these systems, one may never treat macroscopic (or
mesoscopic) motions as pre-fixed
protocols.

There has recently been a revival in attempts to explore the
boundaries of the second law. e.g. in \cite{ev}.  Here we stick to
derivations of the second law in the average, not focusing on
path-dependent fluctuations of work or entropy.  Instead of
presenting a conceptually unifying and general scheme for second
law scenario's and derivations such as in \cite{dmn}, we prove
concrete estimates for a specific set-up which is wide enough to
cover such interesting physics as limitations on the working of
biological and thermodynamic engines and as the impact of
macroscopic bodies.
Discussions of  various related issues and extensions to quantum systems will appear elsewhere \cite{next}.

\paragraph*{Setting ---}
Consider an isolated classical mechanical system consisting of $N$
(microscopic) particles confined in a finite volume. We start from
a completely arbitrary $N$-body Hamiltonian
\begin{equation}
H_\mathrm{tot}=\sum_{j=1}^N\frac{|\tilde{\bsp}_j|^2}{2\mu_j}+U_\mathrm{tot}(\tilde{\bsr}_1,\ldots,\tilde{\bsr}_N),
\label{e:H1}
\end{equation}
where $\tilde{\bsr}_j$, $\tilde{\bsp}_j$, and
  $\mu_j$ are the coordinate, the momentum, and the mass, respectively, of the $j$-th particle.
The potential $U_\mathrm{tot}(\tilde{\bsr}_1,\ldots,\tilde{\bsr}_N)$ is
 arbitrary, and represents both the interaction between the particles and the external forces acting on them.

We imagine that $N$ is a huge number, and we extract $n$ degrees
of freedom, with $n=O(1)$, to describe the macroscopic motions in
the system. We let $I_1,\ldots,I_n\subset\{1,\ldots,N\}$ be
non-overlapping sets of indices (where we do not assume
$\cup_{i=1}^n I_i=\{1,\ldots,N\}$), and denote by $m_i:=\sum_{j\in
I_i}\mu_j$ the mass, by $\bsr_i:=(\sum_{j\in
I_i}\mu_j\tilde{\bsr}_j)/m_i$ the center of mass, and by
$\bsp_i:=\sum_{j\in I_i}\tilde{\bsp}_j$ the momentum,
respectively, of the $i$-th macroscopic body \cite{en:AB}. We
denote these macroscopic coordinates collectively as
$\bsR:=(\bsr_1,\ldots,\bsr_n)$ and $\bsP:=(\bsp_1,\ldots,\bsp_n)$.
We denote the canonical coordinates and momenta of {\em all}\/ the
remaining $3(N-n)$ microscopic degrees of freedom collectively as
$X$. One should think of the macroscopic degrees of freedom
described by $\bsR$, $\bsP$ as our observable and controllable
system, while the $X$ describe the internal and uncontrolled
reservoir-like variables.

Using these new variables, we decompose the total Hamiltonian \eqref{e:H1} as
$H_\mathrm{tot}=H(\bsP,\bsR)+h_\bsR(X)$,
where
\begin{equation}
H(\PR)=\sum_{i=1}^n\frac{|\bsp_i|^2}{2m_i}+U(\bsR),
\label{e:HM}
\end{equation}
is the macroscopic part and  $h_\bsR(X)$  is the remaining
microscopic part of the Hamiltonian.
Although this decomposition is not unique,  our main results do not depend
on the particular decomposition.
Note that  $h_\bsR(X)$ can
depend on the macroscopic position $\bsR$ (but not on $\bsP$)
while $H(\PR)$ does not depend on $X$.

The internal degrees of freedom $X$ are initially in a
restricted thermal equilibrium; the macroscopic degrees of freedom
$(\bsP,\bsR)$ start sharply peaked around some fixed values. More
precisely we sample the initial condition $(\bsP,\bsR,X)$
according to the probability distribution
\begin{equation}
\nu_0(\PRX)=\tilnu_0(\PR)\,e^{-\beta\{h_{\bsR}(X)-F_0(\bsR)\}}\,
\chi[X\in\calC(\bsR)].
\label{e:nu0}
\end{equation}
A priori $\tilnu_0(\PR)$ can be taken as an arbitrary normalized
distribution,
 but we usually take it as a distribution that is peaked around fixed values $\bsP^{(0)}$ and  $\bsR^{(0)}$.
The $X$ are in thermal equilibrium at inverse temperature
$\beta=1/kT$ but with a possible restriction; a typical example
being a gas confined in a certain region. Here $\chi[\cdot]$ is
the indicator defined by $\chi[\text{true}]=1$ and
$\chi[\text{false}]=0$, and $\calC(\bsR)$ is the range of $X$
determined by the initial (macroscopic) setting. The (restricted)
free energy is then $F_0(\bsR)=-\beta^{-1}\log\int
dX\,e^{-\beta\,h_{\bsR}(X)}\,\chi[X\in\calC(\bsR)]$.

After choosing the initial state $(\bsP,\bsR,X)$ according to
\eqref{e:nu0}, the system evolves following the Hamiltonian
dynamics determined by \eqref{e:H1} (and thus possibly leaving the
initial constraint $\calC(\bsR)$). While there is little hope of
solving the fully interacting $N$-particle dynamics exactly, we
can nevertheless derive completely rigorous inequalities about the
energy transfer.

\paragraph*{Main results ---}
Let $\nu_t(\PRX)$ be the probability distribution at time
$t\ge0$ determined uniquely from the initial distribution
 \eqref{e:nu0} and from the Hamiltonian dynamics.
We denote by $\sbkt{\cdots}_t=\int d\bsP d\bsR dX\,\nu_t(\PRX)\,(\cdots)$ the corresponding
 expectation value.
Let $\tilnu_t(\PR):=\int dX\,\nu_t(\PRX)$ be the probability distribution
 at $t$ of the macroscopic degrees of freedom, and
 $S_t:=-\int d\bsP d\bsR\,\tilnu_t(\PR)\,\log\tilnu_t(\PR)$ the corresponding Shannon entropy.
We finally let $F(\bsR)=-\beta^{-1}\log\int
dX\,e^{-\beta\,h_{\bsR}(X)}$ be the free energy without
restriction.

{\em Theorem 1 ---} For any $t\ge0$, one has
\begin{equation}
\sbkt{H}_t-\sbkt{H}_0\le\sbkt{F_0(\bsR)}_0-\sbkt{F(\bsR)}_t+\frac{1}{\beta}(S_t-S_0).
\label{e:main1}
\end{equation}

 A first useful interpretation of that bound may be obtained by
 rewriting \eqref{e:main1} as
 \begin{equation}\label{feb}
 \sbkt{F(\bsR)}_t+(\sbkt{H}_t-S_t/\beta)
 \le
 \sbkt{F_0(\bsR)}_0+(\sbkt{H}_0-S_0/\beta),
 \label{e:Fdec}
 \end{equation}
 implying that the total free energy, when properly defined as above, cannot increase in
 time.
 One can further rewrite \eqref{e:Fdec} in the form of the Clausius inequality \cite{next,MN}.
 Another important interpretation of \eqref{e:main1} will be discussed as the example ii) below.

One expects that the change $S_t-S_0$ of the entropy of the macroscopic part is of $O(1)$ for a suitable choice of the initial distribution.
To show this precisely, we further specify the
situation. Let us take for $\tilnu_0(\PR)$ the uniform
distribution in a very small region of the phase space around a
fixed initial value $(\bsP^{(0)},\bsR^{(0)})$. The volume of the
region in the $\bsP$-space is taken to be
$v_\mathrm{m}=\prod_{i=1}^n (2m_i/\beta)^{3/2}$, which roughly
means that each component of the momenta can fluctuate only of the
order of the thermal momentum $\sqrt{2m_i/\beta}$. The volume of
the region in the $\bsR$-space is $V_\mathrm{c}=(v_0)^n$, where
$v_0$ is a very small spatial volume that corresponds to the
initial fluctuation of the center of mass of the macroscopic
bodies.

Let us fix the final time $t\ge0$. We assume that the final
distribution $\tilnu_t(\PR)$ is nonvanishing only when
$\bsR\in\calR$, where $\calR$ is a finite region with the
($3n$-dimensional) volume $V_\mathrm{c}$. When the whole system is
enclosed in a box of volume $V$, we can take $V_\mathrm{c}=V^n$.
We write $\eta=V_\mathrm{c}/v_\mathrm{c}$. Finally we define
\begin{equation}
K=\beta\,[\mathop{\mathrm{max}'}_{(\PR)}
\{H(\PR)+F_0(\bsR)\}
-\min_{\bsR\in\calR}\{U(\bsR)+F(\bsR)\}\,],
\label{e:K}
\end{equation}
where the maximum is taken over $(\PR)$ such that
$\tilnu_0(\PR)\ne0$. The quantity $K$ may be interpreted as the
maximally allowed kinetic energy of the macroscopic part in the
final state when one assumes that the maximum work principle is
valid.

{\em Theorem 2 ---} For any $t\ge0$, one has
\begin{equation}
\sbkt{H}_t-\sbkt{H}_0\le\sbkt{F_0}_0-\sbkt{F}_t+\frac{1}{\beta}(\log\eta+\bar{S}),
\label{e:main2}
\end{equation}
where $\bar{S}\le(3n/2)\log(K+\log\eta)+c_n$ if $K\ge K_n$, and $\bar{S}\le b_n$ if $K\le K_n$.
Here $K_n$, $c_n$, and $b_n$ are constants which depend only on $n$ \cite{en:const}.

The quantity $\log\eta+\bar{S}$ is indeed of $O(1)$ for normal macroscopic systems \cite{en:est}.
Then we can rewrite \eqref{e:main2} as
\begin{equation}
W\le\sbkt{F_0(\bsR)}_0-\sbkt{F(\bsR)}_t+O(kT),
\label{e:MWP}
\end{equation}
where $W=\sbkt{H(\bsP,\bsR)}_t-\sbkt{H(\bsP,\bsR)}_0$ is the total
work done to the macroscopic part by the internal (microscopic)
part. Since $O(kT)$ is a negligibly small energy for macroscopic
bodies,
 \eqref{e:MWP} implies the {\em maximum work principle}\/, which says
 that the work done on the system (the macroscopic coordinates $(\bsP,\bsR)$) cannot
  exceed the decrease of the free energy of the reservoir, here played by
  the internal degrees of freedom $X$. Note that the term $O(kT)$ while small
   and practically irrelevant on macroscales, cannot be avoided in principle.


\paragraph*{Examples}--- {\em i)~Macroscopic impact:}\/
Suppose that one has $\sbkt{F_0(\bsR)}_0=\sbkt{F(\bsR)}_t$ for the
free energy of the internal degrees of freedom. Then \eqref{e:MWP}
becomes $\sbkt{H(\bsP,\bsR)}_t\lesssim\sbkt{H(\bsP,\bsR)}_0$,
which means that macroscopic or collective motion gets halted; the
energy stored in the macroscopic part flows to the internal
degrees.

 Typical and
important examples are impact between macroscopic bodies. Consider
a system of $n$ macroscopic bodies, each of which consists of a
huge number of ``molecules.'' The microscopic potential
$U_\mathrm{tot}(\tilde{\bsr}_1,\ldots,\tilde{\bsr}_N)$, which we
assume to be translation invariant, describes  binding forces
within each body and interactions between the different bodies
when they come close to each other.

We assume that, initially (i.e., before the
collision) and finally (i.e., after the collision), all the bodies are
sufficiently apart from each other so that we have $U(\bsR)=0$ with probability one.
We also assume that initially the internal degrees of freedom within each body are in their unrestricted equilibrium.
Thus we have $\sbkt{F_0(\bsR)}_0=\sbkt{F(\bsR)}_t$, and  \eqref{e:MWP} implies
\begin{equation}
\sum_{i=i}^n\bkt{\frac{|\bsp_i|^2}{2m_i}}_t-\sum_{i=i}^n\bkt{\frac{|\bsp_i|^2}{2m_i}}_0\le O(kT),
\label{e:rest}
\end{equation}
where $O(kT)$ is determined by
$K=\beta\sum_{i=1}^n|\bsp_i^{(0)}|^2/(2m_i)$ for $\bsp_i^{(0)}$ the
initial momentum. In the simplest case of an impact between two
bodies, that implies the (well-known) fact that the coefficient of
restitution does not exceed unity, proved before in a stronger
form but in a special setting with a high symmetry \cite{Hal}.
Here the fact is proved in great generality \cite{en:rot}.

{\em ii)~Thermodynamic machines:}\/
We can treat various ``machines'' whose initial state stores extra
free energy in such forms as  a difference in pressure or chemical
potential.
Then the extra free energy may be converted into mechanical energy by
relaxing the constraints.
Our bound \eqref{e:MWP}
states that the extracted mechanical work can never exceed the
difference between the initial free energy and the final
equilibrium free energy.

{\em iii)~Molecular motor:}\/
If the macroscopic part settles at time $t$ to a configurationally equivalent state as the initial state,
we can assume $S_0=S_t$. Then \eqref{e:main1} yields the 
maximum work principle $W\le\sbkt{F_0(\bsR)}_0-\sbkt{F(\bsR)}_t$
without a small correction.

The above assumption is justified in the  so called steady state
regime, in which the macroscopic part makes a successive
transition between quasi stationary states. An important example
is the rotational motion in a molecular motor such as the F${}_1$
ATP synthase \cite{F1}. We identify the rotational degrees of
freedom of the motor as the single macroscopic coordinate,
and couple it with an external potential (provided, for example,
by an elastic filament attached to the ``rotating shaft'' of the
motor). We choose the initial state as the moment where an ATP is
captured by the motor protein, and assume that the ATP hydrolysis
proceeds as a semi-mechanical process synchronized with the
rotational motion. Then we may assume
$\sbkt{F_0}_0-\sbkt{F}_t$  is equal to $\Delta G_{\rm
ATP}\simeq20kT$, which is the free energy released in the hydrolysis of
an ATP. Our bound \eqref{e:main1} implies
$\sbkt{U(\bsr_1)}_t-\sbkt{U(\bsr_1)}_0\le\Delta G_{\rm ATP}$,
which guarantees that the second law is valid {\em in average}.
Obviously, for such small systems like molecular motors,
fluctuations may be more important than in usual thermodynamics
for large systems.

\paragraph*{Proof of Theorem 1 ---}
We fix the final time $t\ge0$ throughout the proof.
Let us denote the initial coordinates collectively as $\Gamma=(\PRX)$.
It is convenient to denote the state at time $t$ by different variables as $\Gamma'=(\PRXp)$.
There is a one-to-one correspondence between $\Gamma$ and $\Gamma'$ determined by the time-evolution.
When we regard one of them as a function of the other, we explicitly write $\Gamma(\Gamma')$ or $\Gamma'(\Gamma)$.
We also freely use notations like $\bsP(\Gamma')$ or $X'(\Gamma)$.

Let us consider a probability distribution at time $t$
\begin{equation}
\mu_t(\Gamma')=\tilmu_t(\PRp)\,e^{-\beta\{h_{\bsR'}(X')-F(\bsR')\}},
\label{e:mut}
\end{equation}
where $\tilmu_t(\PRp)$ is a normalized distribution which will be chosen later.
We denote by $\mu_0(\Gamma)$ the probability distribution at the initial time obtained from the inverse time evolution.
We follow the idea in \cite{MN}, and examine the relative entropy
\begin{eqnarray}
&&D(\mu_0|\nu_0):=
\int d\Gamma\,\nu_0(\Gamma)\,\log\frac{\nu_0(\Gamma)}{\mu_0(\Gamma)}
\ret
&&
=\int d\Gamma\,\nu_0(\Gamma)\,\log\nu_0(\Gamma)
-\int d\Gamma'\,\nu_t(\Gamma')\,\log\mu_t(\Gamma'),
\label{e:Dmn1}
\end{eqnarray}
where the final expression follows from the Liouville theorem.
By using the explicit form \eqref{e:nu0} and \eqref{e:mut}, we get
\begin{eqnarray}
&&D(\mu_0|\nu_0) =\int d\Gamma\,\nu_0(\Gamma)\bigl[
\log\tilnu_0(\PR)-\beta\{h_\bsR(X)-F_0(\bsR)\}\bigr]
\ret
&& -\int
d\Gamma'\,\nu_t(\Gamma')\bigl[
\log\tilmu_t(\PRp)-\beta\{h_{\bsR'}(X')-F(\bsR')\}\bigr] \ret
&&= -S_0-\beta\sbkt{h_\bsR(X)}_0+\beta\sbkt{F_0(\bsR)}_0
\ret
&&
+S(\tilnu_t,\tilmu_t)+\beta\sbkt{h_{\bsR'}(X')}_t-\beta\sbkt{F(\bsR')}_t,
\label{e:Dmn2}
\end{eqnarray}
where we have defined
\begin{equation}
S(\tilnu_t,\tilmu_t)=-\int
d\bsP'd\bsR'\,\tilnu_t(\PRp)\,\log\tilmu_t(\PRp). \label{e:Snm}
\end{equation}
By using the energy conservation
law $\sbkt{H(\PR)+h_{\bsR}(X)}_0=\sbkt{H(\PRp)+h_{\bsR'}(X')}_t$ and the fact that
the relative entropy $D(\mu_0|\nu_0)$ is always nonnegative, we get our basic inequality
\begin{eqnarray}
&&\sbkt{H(\PRp)}_t-\sbkt{H(\PR)}_0
\ret&&\le
\sbkt{F_0(\bsR)}_0-\sbkt{F(\bsR')}_t
+\frac{1}{\beta}\{S(\tilnu_t,\tilmu_t)-S_0\}. \label{e:themain}
\end{eqnarray}
By choosing $\tilmu_t(\PRp)=\tilnu_t(\PRp)$ (which is indeed the best choice),
 \eqref{e:themain} reduces to the desired \eqref{e:main1} which ends the proof of Theorem 1.

\paragraph*{Proof of Theorem 2 ---}
We first control the  distribution $\hat{\nu}_t(\bsP'):=\int d\bsR' dX'\nu_t(\Gamma')$.
We again use the Liouville theorem, the energy conservation, and \eqref{e:nu0} to get
\begin{eqnarray}
&&\hat{\nu}_t(\bsP')=\int d\bsR' dX'\nu_0(\Gamma(\Gamma'))
\ret
&&=\int d\bsR' dX'\,\tilnu_0(\bsP(\Gamma'),\bsR(\Gamma'))\,\chi[X(\Gamma')\in\calC(\bsR(\Gamma'))]\times
\ret
&&
\times\exp[\beta\{-h_{\bsR'}(X')-H(\PRp)
\ret
&&\hspace{1.5cm}
+H(\bsP(\Gamma'),\bsR(\Gamma'))+F_0(\bsR(\Gamma'))\}]
\label{e:nutP}
\end{eqnarray}
We can bound that from above by replacing
$H(\bsP(\Gamma'),\bsR(\Gamma'))$ and $F_0(\bsR(\Gamma'))$ by
$H_0^\mathrm{max}=\mathrm{max'}_{(\PR)}H(\PR)$ and
$F_0^\mathrm{max}=\mathrm{max'}_{(\PR)}F_0(\bsR)$ (where the max
are taken over $(\PR)$ such that $\tilnu_0(\PR)\ne0$),
respectively, and $H(\PRp)$ by
$\sum_{i=1}^n|\bsp'_i|^2/(2m_i)+U^\mathrm{min}$ where
$U^\mathrm{min}=\min_{\bsR'\in\calR}U(\bsR')$. After these
replacements, we can perform the $X'$ integral to get
$\int dX'\,e^{-\beta\,h_{\bsR'}(X')}=e^{-\beta\,F(\bsR')}\le e^{-\beta\,F^\mathrm{min}}$
with $F^\mathrm{min}=\min_{\bsR'\in\calR}F(\bsR')$. We finally use
the bound $\tilnu_0(\PR)\le(v_\mathrm{c}\,v_\mathrm{m})^{-1}$, and
then integrate over $\bsR'$ to get the total volume
$V_\mathrm{c}$. The final result reads
\begin{equation}
\tilnu_t(\bsP')\le\frac{\eta}{v_\mathrm{m}}\,
\exp\bigl[-\beta\sum_{i=1}^n\frac{|\bsp'_i|^2}{2m_i}+K\bigr],
\label{e:tilnu}
\end{equation}
with
$K:=\beta\{H_0^\mathrm{max}+F_0^\mathrm{max}-U^\mathrm{min}-F^\mathrm{min}\}$
which is the same as \eqref{e:K}.
The bound \eqref{e:tilnu} shows that the probability
to find large final kinetic energy decays exponentially for
sufficiently large energy; the bound will prove to be essential
for our final estimate.

We use \eqref{e:tilnu} to bound $S(\tilnu_t,\tilmu_t)$. Since
$\tilmu_t(\PRp)$ is arbitrary, we choose it as
$\tilmu_t(\PRp)=\hat{\mu}(\bsP')\,(V_\mathrm{c})^{-1}\,\chi[\bsR'\in\calR]$,
where $\hat{\mu}(\bsP')$ will be chosen later. Then we get
$S(\tilnu_t,\tilmu_t)=\log V_\mathrm{c}+\hat{S}$ with
$\hat{S}=-\int d\bsP'\,\hat{\nu}(\bsP')\,\log\hat{\mu}(\bsP')$. To
bound $\hat{S}$, we make a change of variables. Let
$\bst_i=\sqrt{\beta/(2m_i)}\,\bsp'_i$, and
$\bst=(\bst_1,\ldots,\bst_n)\in\mathbf{R}^{3n}$. Correspondingly
we set $\bar{\nu}(\bst):=v_\mathrm{m}\,\hat{\nu}(\bsP')$ and
$\bar{\mu}(\bst):=v_\mathrm{m}\,\hat{\mu}(\bsP')$ with (as before)
$v_\mathrm{m}=\prod_{i=1}^n(2m_i/\beta)^{3/2}$. Then we have
$\hat{S}=\log v_\mathrm{m}+\bar{S}$ with $\bar{S}=-\int
d\bst\,\bar{\nu}(\bst)\,\log\bar{\mu}(\bst)$. Since
$S_0=\log(v_\mathrm{c}v_\mathrm{m})$ for the present initial
distribution, the basic inequality \eqref{e:themain} reduces to
\eqref{e:main2}.

The remaining task is to bound $\bar{S}$ by using \eqref{e:tilnu}:
$\bar{\nu}(\bst)\le\exp(-|\bst|^2+K')$ with $K'=K+\log\eta$. That
is an elementary technical problem, which can be solved in various
manners. Let us present a simple bound. Take a constant $A>0$ such
that $v_A:=\int_{|\bst|_\infty\ge
A}d\bst\,\exp[-\exp[|\bst|^2/2]]\le1/2$ (where
$|\bst|_\infty=\max_{i=1}^{3n}|t_i|$). Then we set
$\bar{\mu}(\bst)=\exp[-\exp[|\bst|^2/2]]$ for $|\bst|_\infty\ge
A$, and  $\bar{\mu}(\bst)=(2A)^{-3n}(1-v_A)\ge\{2(2A)^{3n}\}^{-1}$
for $|\bst|_\infty<A$, satisfying $\int d\bst\,\bar{\mu}(\bst)=1$.
With that choice of $\bar{\mu}(\bst)$, we bound $\bar{S}$ as
\begin{equation}
\bar{S}\le\log\{2(2A)^{3n}\}+\int_{|\bst|_\infty\ge A}d\bst\,e^{-|\bst|^2/2+K'}.
\label{e:barS}
\end{equation}
The integral in the second term is bounded as
\begin{eqnarray}
&&\int_{|\bst|_\infty\ge A}d\bst\,e^{-|\bst|^2/2+K'}\le
6ne^{K'}\int_{t_1\ge A}d\bst\,e^{-|\bst|^2/2} \ret
&&=
6ne^{K'}\int_A^\infty dt\,e^{-t^2/2}(\int_{-\infty}^\infty
dt\,e^{-t^2/2})^{3n-1}
\ret&&
\le 3n(2\pi)^{3n/2}e^{K'-A^2/2},
\label{e:int}
\end{eqnarray}
where we noted that $\int_A^\infty dt\,e^{-t^2/2}=\int_0^\infty
ds\,e^{-(s+A)^2/2}\le e^{-A^2/2}\int_0^\infty ds\,e^{-s^2/2}$. We
 set $A=\sqrt{2K'}$ (provided that this value is large enough
to guarantee $v_A\leq 1/2$) to get
$\bar{S}\le\log (K')^{3n/2}+\log\{2(2\sqrt{2})^{3n}\}+3n(2\pi)^{3n/2}$.
When $K'$ is not large enough, we choose $A$ to be the minimum allowed value.

It is a pleasure to thank Hisao Hayakawa, Christopher Jarzynski,
Tadas Nakamura, Karel Neto\v cn\'y, Takayuki Nishizaka and
Shin-ichi Sasa for valuable discussions.


\begin{thebibliography}{10}
\bibitem{en:work}
A formulation of the second law that comes close to the maximum
work principle is also referred to as Kelvin's formulation, that
it is impossible for a system operating in a cycle and in contact
with one thermal reservoir to perform positive work in the
surroundings.
\bibitem{en:AB}
We write $A:=B$ when $A$ is defined in terms of $B$.
\bibitem{en:const}
A possible choice is $K_n=\log\{6n(2\pi)^{3/2}\}$,
$c_n=\log\{2(2\sqrt{2})^{3n}\}+3n(2\pi)^{3n/2}$, and
$b_n=\log\{2(2\sqrt{2K_n})^{3n}\}+e^{K_n}/2$. These constants may
be improved if necessary.
\bibitem{en:est}
Let us set, for example, $\eta\sim(1\,\mathrm{m}/0.1\,\mathrm{mm})^{3n}= 10^{12n}$ and
$K\sim n\,(k\,300\,\mathrm{K})^{-1}(1\,\mathrm{J})\sim n\,10^{20}$.
Even with these  rather radical overestimates, we have
$\log\eta+\bar{S}\sim100 n$. It is needless to say that $100kT$ (which is
roughly $3\times 10^{-19}\,\mathrm{J}$ at $T=300\,\mathrm{K}$) is
a ridiculously small energy for a macroscopic degree of freedom.
\bibitem{en:rot}
One can also treat the effect of macroscopic rotation  \cite{next}.
\end{thebibliography}
\end{document}